\documentclass[12pt,pra,aps,amssymb,amsfonts,amsmath,tightenlines]{revtex4}

\usepackage{dsfont}
\usepackage{graphicx}
\usepackage{amssymb,amsfonts, bbm}

\newtheorem{Proposition}{Proposition}
\newtheorem{Lemma}{Lemma}

\newcommand{\half}{\mbox{$\textstyle \frac{1}{2}$}}

\newcommand{\re}{\mbox{$\rm e$}}
\newcommand{\rd}{\mbox{$\rm d$}}
\begin{document}

\title{L\'evy-Vasicek Models and the Long-Bond Return Process}
\author{Dorje C. Brody${}^{1,2}$, Lane P. Hughston${}^{1}$, and David M. Meier${}^{1}$}

\affiliation{${}^{1}$Department of Mathematics, Brunel University London, Uxbridge UB8 3PH, UK \\ 
${}^{2}$St Petersburg National Research University of Information Technologies, Mechanics and Optics,
49 Kronverksky Avenue, St Petersburg 197101, Russia}

\date{\today}

\begin{abstract}
\noindent 
The classical derivation of the well-known Vasicek model for interest rates is reformulated in terms of the associated pricing kernel. An advantage of the pricing kernel method is that it allows one to generalize the construction to the L\'evy-Vasicek case, avoiding issues of market incompleteness. In the L\'evy-Vasicek model the short rate is taken in the real-world measure to be a mean-reverting process with a general one-dimensional L\'evy driver admitting exponential moments. Expressions are obtained for the L\'evy-Vasicek bond prices and interest rates, along with a formula for the return on a unit investment in the long bond, defined by $L_t = \lim_{T \rightarrow \infty} P_{tT} / P_{0T}$, where $P_{tT}$ is the price at time $t$ of a $T$-maturity discount bond. We show that the pricing kernel of a L\'evy-Vasicek model is uniformly integrable if and only if the long rate of interest is strictly positive.
\end{abstract}

\maketitle

\section{Pricing kernels}
\label{Pricing kernels} 

\noindent
The Vasicek model (Vasicek 1977) is one of the oldest and most well-studied models in the mathematical finance literature, and one might think that there is very little that is new that can be said about it. But it turns out that there are some surprising features of the Vasicek model relating to the long rate of interest that are very suggestive when it comes to modelling long term interest rates in general.

We fix a probability space $(\Omega, \mathcal{F}, \mathbb{P})$ with filtration $\{{\mathcal{F}_t}\}_{t\geq0}$. Time $0$ denotes the present. The probability measure $\mathbb{P}$ is the physical measure, and $\{{\mathcal{F}_t}\}$ represents the flow of market information. We introduce an appropriate unit of account, and for $T$ and $t$ such that $0 \leq t < T$ we let $P_{tT}$ denote the value at time $t$ of a discount bond that pays out one unit of account at maturity $T$. 
In what follows we use a pricing kernel method to construct the Vasicek model. This is not the way in which the Vasicek model is usually presented in the literature. Nevertheless, the pricing kernel approach is very effective. In particular, the pricing kernel formulation of the classical Vasicek model leads us directly to a construction of the corresponding L\'evy-Vasicek model, extending results of Cairns (1999), Eberlein \& Raible (1999), Norberg (2004), and others.  

We begin with a few remarks about pricing kernels and then turn to the case of the Vasicek model. Let us recall briefly how pricing kernels work in the elementary geometric Brownian motion (GBM) model for asset prices. We fix a Brownian motion $\{W_t\}_{t\geq0}$ on $(\Omega,{\mathcal F},{\mathbb P})$, and take it to be adapted to $\{{\mathcal F}_t\}$. The GBM model is characterized by the specification of a pricing kernel along with a collection of one or more so-called investment-grade assets. 
We assume for simplicity that the assets pay no dividends over the time horizon considered. The idea of an investment-grade asset is that it should offer a positive excess rate of return above the interest rate. Ordinary stocks and bonds are in this sense investment-grade, whereas put options are not. 

For the pricing kernel in the GBM model we assume that we have an expression of the form%
\begin{equation}
\pi_t = \re^{-rt}\, \re^{-\lambda W_t - \frac{1}{2}\lambda^2t}, \label{GBMPK}
\end{equation} 
where $r$ is the interest rate, and $\lambda >0$ is a risk aversion parameter. We require that the product of the pricing kernel and the asset price should be a ${\mathbb P}$-martingale. Let us suppose therefore that for some $\beta \geq -\lambda$ the product takes the form
\begin{eqnarray}
\pi_t S_t = S_0 \, \re^{\beta W_t - \frac{1}{2}\beta^2t},
\end{eqnarray}
where $S_t$ denotes the value of the asset at time $t$. For a typical investment-grade asset in the GBM model we thus have
\begin{eqnarray}
S_t = S_0 \, \re^{(r+\lambda\sigma)t} \,  \re^{\sigma W_t - \frac{1}{2}\sigma^2t}, 
\end{eqnarray}
where $\sigma = \beta + \lambda$. 
The term $\lambda\sigma$ is called the risk premium or excess rate of return, and it is positive under 
the assumptions we have made. The idea of a ``pricing kernel" dates back to the 1970s, and is used for example by Ross (1978). The alternative term ``market kernel" is used by Garman (1976). Authors have employed a variety of terms for essentially the same concept. Economists speak of the ``marginal rate of substitution". The term ``state price density" appears in Dothan \& Williams (1978). The term ``stochastic discount factor" is used in Cox \& Martin (1983). The term ``state price deflator" is used in Duffie (1992).  

Pricing kernel models are discussed in detail in, for example, Cochrane (2005) and Hunt \& Kennedy  (2004). If the risky asset under consideration is a European-style derivative whose terminal payoff is $H_T$, then the value of the derivative at time $t<T$ is given by 
\begin{eqnarray} 
H_t = \frac{1}{\pi_t} {\mathbb E}_t[\pi_T H_T]. 
\end{eqnarray} 
In particular, if the derivative pays one unit of account so that $H_T=1$, then we recover the pricing formula for a discount bond, given by
\begin{eqnarray}
P_{tT} = \frac{ 1} {\pi_t} \, \mathbb{E}_t[\pi_T].  
\label{discount bond formula} 
\end{eqnarray}
We refer to the process $\{ n_t\}_{t \geq 0}$ defined by $n_t = 1/\pi_t$ as the ``natural numeraire" (Flesaker \& Hughston 1997), or ``growth-optimal portfolio". It serves as a benchmark, relative to which other non-dividend-paying assets are martingales. As an example of derivative pricing in the GBM model, one can consider the valuation of a digital put on the natural numeraire, with unit notional, strike $\kappa$,  and maturity $T$. In this case we have
\begin{equation}
	H_T = \mathds{1}\left\{ n_T < \kappa \right\},
\end{equation}
where $\mathds{1}\{ \, \cdot \, \}$ denotes the indicator function. Using the pricing kernel  \eqref{GBMPK},  a straightforward calculation gives
\begin{equation}
	H_0 = \re^{-rT} N\left[ \frac{ \log \left( \re^{-rT} \kappa \right) + \half \lambda^2 T}{\lambda \, T^{1/2}}\right],
\end{equation}
where $N[\, \cdot \,]$ is the normal distribution function. We mention the example of a digital put on the natural numeraire because it turns out to be relevant later in our consideration of the uniform integrability of the pricing kernel. 
\section{Vasicek pricing kernel}
\label{Vasicek pricing kernel}

\noindent
We shall extend the geometric Brownian motion model by keeping the risk-aversion level 
constant, but allowing the interest rate to be stochastic. The pricing kernel can then be expressed 
in the form 
\begin{eqnarray}
\pi_t = \exp \left[-\int_0^t r_s {\rm d}s  -\lambda W_t - \frac{1}{2}\lambda^2t \right]. 
\label{Vasicek pricing kernel}
\end{eqnarray} 
In the Vasicek model, the short rate process $\{r_t\}_{t \geq 0}$ is taken to be a mean-reverting 
process of the Ornstein-Uhlenbeck (OU) type, satisfying 
\begin{eqnarray}
\rd r_t = k(\theta - r_t) \rd t - \sigma \rd W_t. 
\label{OU dynamics}
\end{eqnarray} 
Here $k$, $\theta$, and $\sigma$ denote the mean reversion rate, the mean reversion level, and the absolute volatility of the short rate. We choose $\sigma$ to be positive. The minus sign in front of $\sigma$ in the dynamics of the interest rate is a convention that ensures that the discount bond volatility is positive. Note that the volatility parameter here has units of $T^{-3/2}$, in contrast with the $\sigma$ of the GBM model, which has units of $T^{-1/2}$. The initial value of the interest rate is $r_0$. The dynamical equation (\ref{OU dynamics}) can then be solved by use of an integrating 
factor to give 
\begin{eqnarray}
r_t = \theta + (r_0 - \theta)\, \re^{-kt} -\sigma \int_0^t \re^{k(s- t)} \rd W_s. 
\label{short rate}
\end{eqnarray}
To determine the pricing kernel we require an expression for the integrated 
short rate, 
\begin{eqnarray}
I_t = \int_0^t r_s \rd s .
\label{integrated short rate}
\end{eqnarray}
Substitution of (\ref{short rate}) into (\ref{integrated short rate}) gives 
\begin{eqnarray}
I_t = \theta t + \frac{1}{k} \left(1 - {\rm e}^{-kt}\right) (r_0 - \theta) - \sigma \int_{s=0}^t \int_{u=0}^s \re^{k(u - s)} \rd W_u \rd s. 
\end{eqnarray}
The double integral can be rearranged according to the scheme 
\begin{eqnarray}
\int_{s=0}^t \int_{u=0}^s \re^{k(u - s)} \rd W_u \rd s =
\int_{u=0}^t \int_{s=u}^t \re^{k(u - s)} \rd s \rd W_u =\frac{1}{k} \int_0^t (1 - \re^{k(u-t)}) \rd W_u,
\end{eqnarray}
from which it follows that
\begin{eqnarray}
I_t =  \theta t + \frac{1}{k} \left(1 - {\rm e}^{-kt}\right)  (r_0 - \theta) - \frac{\sigma}{k} \int_0^t (1 - \re^{k(u-t)}) \rd W_u. 
\label{integrated short rate formula}
\end{eqnarray}
On account of (\ref{short rate}) we can then replace the stochastic integral above with an expression involving the short rate to 
obtain 
\begin{eqnarray} 
I_t = \theta t + \frac{1}{k} (r_0 - r_t) -\frac{\sigma}{k} W_t \,.  
\end{eqnarray}
It follows that the Vasicek pricing kernel can be expressed in the form 
\begin{eqnarray} 
\pi_t =  \exp \left[ - \left(\theta+ \frac{1}{2}\lambda^2\right) t   + \left(\frac{\sigma}{k} - \lambda\right) 
W_t  -  \frac{1}{k} (r_0 - r_t) \right] . 
\label{Vasicek pricing kernel with short rate} 
\end{eqnarray}

Note the appearance of the 
``naked" $W_t$ in the formula for the pricing kernel. Often it is said that the Vasicek model has a single state variable,  the short rate. This statement is a little misleading. For while it is true, as we shall see shortly, that the price at time $t$ of a $T$-maturity discount bond depends only on the state variable $r_t$ insofar as its stochasticity is concerned, the pricing kernel depends at time $t$ on a pair of state variables, namely, $r_t$ and $W_t$. Likewise, the value of a unit-initialized money market account
\begin{eqnarray} 
B_t = \exp \int_0^t r_s \rd s 
\end{eqnarray}
depends on both $r_t$ and $W_t$. We take the view that to specify a financial model one needs to give the price processes for the basic assets of the model, together with the pricing kernel process. In the case of an interest rate model this means giving the processes for discount bonds of all maturities, the money market account, and the pricing kernel. Thus, the pricing kernel is part of the model, not a secondary object that one works out in some way from the specification of the processes of basic assets.

An alternative expression for the Vasicek pricing kernel, also useful for our purposes, is obtained by substituting (\ref{integrated short rate formula}) into (\ref{Vasicek pricing kernel}). Then we have
\begin{equation}
 \pi_t = \exp  \left[  -\left(\theta + \frac{1}{2} \lambda^2\right) t - \frac{1}{k} \left(1 - {\rm e}^{-kt}\right)(r_0 - \theta) + \int_0^t \left( \frac{\sigma}{k} - \lambda - \frac{\sigma}{k} {\re}^{k(u-t)} \right) \rd W_u \right].
 \label{explicit pricing kernel}
\end{equation}
In this case, it seems that we have succeeded in specifying the pricing kernel in terms of a single state variable, namely, the value of the stochastic integral appearing in the third term on the right side of  (\ref{explicit pricing kernel}). This is indeed so, but it is distinct from the short rate, so the model as a whole requires two state variables. 
\section{Discount bonds}
\label{Derivation of the discount bond formula}
\noindent
We proceed to derive an expression for $P_{tT}$. In the derivation we find it convenient to work with logarithms rather than exponentials. Thus, instead of  (\ref{explicit pricing kernel}) we write
\begin{equation}
  \log \pi_t = -\left(\theta + \frac{1}{2} \lambda^2\right) t - \frac{1}{k} \left(1 - {\rm e}^{-kt}\right)(r_0 - \theta) + \int_0^t \left( \frac{\sigma}{k} - \lambda - \frac{\sigma}{k} {\rm e}^{k(u-t)} \right) \rd W_u. \label{PK_B_case}
\end{equation}
It follows that
\begin{align}
  \log \pi_T &= -\left(\theta + \frac{1}{2} \lambda^2\right) T - \frac{1}{k} \left( 1- {\rm e}^{-kT}\right) (r_0 - \theta) + 
\int_0^t \left( \frac{\sigma}{k} - \lambda - \frac{\sigma}{k} {\rm e}^{k(u-T)} \right) \rd W_u 
\nonumber \\ 
&\quad + \int_t^T \left( \frac{\sigma}{k} - \lambda - \frac{\sigma}{k} {\rm e}^{k(u-T)} \right) \rd W_u,
\end{align}
and hence for $t < T$ we have
\begin{align}
  \mathbb{E}_t [\pi_T] &= \exp \left[  -\left(\theta + \frac{1}{2} \lambda^2\right) T - \frac{1}{k} \left( 1- {\rm e}^{-kT}\right) (r_0 - \theta) + 
\int_0^t \left( \frac{\sigma}{k} - \lambda - \frac{\sigma}{k} {\rm e}^{k(u-T)} \right) \rd W_u \right]  \nonumber \\
&\quad \quad \times \mathbb{E}_t \exp\left[ \int_t^T \left( \frac{\sigma}{k} - \lambda - \frac{\sigma}{k} {\rm e}^{k(u-T)} \right) \rd W_u \right].
\end{align}
It is a standard result that for any Borel function $\{\alpha_t\}$ satisfying
\begin{align}
\int_t^T \alpha_u^2 \rd u < \infty
\end{align}
we have
\begin{align}
  \mathbb{E}_t  \left[ \exp \left( \int_t^T \alpha_u \rd W_u \right) \right] = \exp \left( \frac{1}{2} \int_t^T \alpha_u^2 \rd u\right).
\end{align}
As a consequence, we obtain
\begin{align}
  \mathbb{E}_t \exp\left[ \int_t^T \left( \frac{\sigma}{k} - \lambda - \frac{\sigma}{k} {\rm e}^{k(u-T)} \right) \rd W_u \right] = \exp \left[ \frac{1}{2} \int_t^T \left( \frac{\sigma}{k} - \lambda - \frac{\sigma}{k} {\rm e}^{k(u-T)} \right)^2 \rd u\right]. 
\end{align}
Therefore, by (\ref{discount bond formula}) we have
\begin{align}
  \log P_{tT} &=  -\left( \theta + \frac{1}{2} \lambda^2\right) (T - t) - \frac{1}{k} \left( {\rm e}^{-kt} - {\rm e}^{-kT}\right)(r_0 - \theta) \nonumber \\
&\quad + \frac{1}{2}  \int_t^T \left( \frac{\sigma}{k} - \lambda - \frac{\sigma}{k} {\rm e}^{k(u-T)} \right)^2 \rd u + \frac{\sigma}{k} \left(1 - {\rm e}^{k(t-T)}\right) \int_0^t {\rm e}^{k(u-t)} \rd W_u. \label{DB_BM}
\end{align}
By use of (\ref{short rate}), in the final term above we can write
\begin{equation}
  \sigma \int_0^t {\rm e}^{k(u-t)} \rd W_u = \theta + (r_0 - \theta) {\rm e}^{-kt} - r_t.
\end{equation}
Then the terms in \eqref{DB_BM} involving $r_0 - \theta$ cancel, and we are left with the following: 
\begin{align}
 \log P_{tT} = -\left( \theta + \frac{1}{2} \lambda^2\right) (T - t) 
 + \frac{1}{2}  \int_t^T \left( \frac{\sigma}{k} - \lambda - \frac{\sigma}{k} {\rm e}^{k(u-T)} \right)^2 \rd u + \frac{1}{k} \left(1 - {\rm e}^{k(t-T)}\right) \left(\theta  - r_t\right).
 \label{log bond price}
\end{align}
Thus, we have isolated the dependence of $P_{tT}$ on the state variable $r_t$. Next, to work out the middle term on the right we note that for $a, b$ constant we have
\begin{equation}
  \int_t^T \left( a- b{\rm e}^{ku}\right)^2 \rd u = a^2(T - t) - \frac{2ab}{k} \left({\rm e}^{kT} - {\rm e}^{kt} \right) + \frac{1}{2} \frac{b^2}{k} \left( {\rm e}^{2kT} - {\rm e}^{2kt}\right). 
\end{equation}
Hence, 
\begin{align}
   \int_t^T \left( \frac{\sigma}{k} - \lambda - \frac{\sigma}{k} {\rm e}^{k(u-T)} \right)^2 \rd u &= \left(\frac{\sigma}{k} - \lambda\right)^2 (T - t) - 2\left( \frac{\sigma^2}{k^3} - \frac{\lambda \sigma}{k^2}\right) \left(1 - {\rm e}^{k(t-T)}\right) \nonumber \\
&\quad \quad+ \frac{1}{2} \frac{\sigma^2}{k^3} \left( 1 - {\rm e}^{2k(t-T)} \right).
\end{align}
Inserting this expression into (\ref{log bond price}), we see that the terms involving $\lambda^2$ cancel, and after some simplification we obtain the following expression for the value of a $T$-maturity discount bond: 
\begin{align}
P_{tT} = \exp \left[  -R_{\infty} (T - t) + \frac{1}{k} \left( 1 - {\rm e}^{k(t-T)} \right) (R_{\infty} - r_t) - \frac{1}{4} \frac{\sigma^2}{k^3}\left( 1- {\rm e}^{k(t-T)}\right)^2 \right] , 
\label{discount bond price}
\end{align}
where
\begin{align}
  R_{\infty} = \theta + \frac{\lambda \sigma}{k} - \frac{1}{2} \frac{\sigma^2}{k^2}.
 \label{Vasicek long rate}
\end{align}
The significance of $R_{\infty}$ is that it represents  
the asymptotic bond yield, or exponential long rate of interest, defined by 
\begin{eqnarray}
R_\infty =  - \lim_{T \to \infty} \frac{1}{T-t} \log P_{tT}.
\end{eqnarray}
 The fact that $R_\infty$ does not depend on $t$ is characteristic of interest rate models for which the tail of the discount function is exponential, and can be viewed as a manifestation of the so-called DIR theorem (Dybvig, Ingersoll \& Ross 1996,
Hubalek, Klein \& Teichmann 2002, Goldammer \& Schmock 2012,  Kardaras \& Platen 2012, Brody \& Hughston 2016). 

\section{Uniform integrability of pricing kernel}

\noindent We remark on a curious feature of the Vasicek 
model that apparently has not been noted previously, namely, that 
$R_\infty>0$ if and only if the pricing kernel is uniformly integrable.  
Before 
we establish this result and its generalization to the L\'evy-Vasicek model, we discuss some financial aspects of the uniform integrability of the pricing kernel. 

We recall (Williams 1991) that a collection $\mathcal {C}$ of random variables is said to be uniformly integrable (UI) if for every $\epsilon>0$ there exists a $\delta \geq 0$ such that for all $X \in \mathcal{C}$ we have
\begin{eqnarray}
{\mathbb E}[|X| {\mathds 1}\{|X|>\delta\}] < \epsilon.
\label{UI property_disp}
\end{eqnarray} 
An equivalent way of expressing the UI condition on $\mathcal {C}$ is
then
\begin{eqnarray}
\lim_{\delta \to \infty} \sup_{X \in \mathcal {C}} {\mathbb E}[|X| {\mathds 1}\{|X|>\delta\}] = 0.
\label{UI property}
\end{eqnarray} 
The limit on the left side of \eqref{UI property} exists since $\sup_{X \in \mathcal {C}} {\mathbb E}[|X| {\mathds 1}\{|X|>\delta\}]$ is decreasing in $\delta$ and bounded from below by zero. 
A random process $\{X_t\}_{t\geq 0}$ is thus said to be UI if for every $\epsilon>0$ there is 
a $\delta>0$ such that 
$
{\mathbb E}[|X_t| {\mathds 1}\{|X_t|>\delta\}] < \epsilon
$
for all $t\geq0$, or equivalently 
\begin{eqnarray}
\lim_{\delta\to\infty} \sup_{t} {\mathbb E}[|X_t| \, {\mathds 1}\{|X_t|>\delta\}] =0 . 
\end{eqnarray} 
For a pricing kernel we can drop the absolute value sign, and the UI condition is that for every $\epsilon>0$ there should exist a $\delta \geq 0$ such that 
$
{\mathbb E}[\pi_t {\mathds 1}\{\pi_t>\delta\}] < \epsilon
\label{PK UI} 
$
for all $t\geq0$, or 
\begin{eqnarray}
\lim_{\delta\to\infty} \sup_{t} {\mathbb E}[\pi_t \, {\mathds 1}\{\pi_t>\delta\}] =0 . \label{PK_UI}
\end{eqnarray}  
Alternatively, the UI condition can be imposed by requiring that for every $\epsilon>0$ there should exist a $\kappa >0$ such that 
\begin{eqnarray}
{\mathbb E}[ \pi_t \, {\mathds 1}\{n_t < \kappa \}] < \epsilon
\label{UI condition} 
\end{eqnarray} 
for all $t\geq0$. Here again we have introduced the natural numeraire $\{n_t\}_{t \geq 0}$ defined by $n_t=1/\pi_t$, and we have set $\kappa =1/\delta$. 
But the left side of (\ref{UI condition}) is the price at time $0$ of a European-style digital put option on the natural
numeraire with strike $\kappa$ and maturity $t$. Thus we have the following:

\begin{Proposition}
A pricing kernel is uniformly 
integrable if and only if for any price level $\epsilon > 0$ there exists a strike $\kappa > 0$ such that the value 
of a digital put option on the natural numeraire is less than $\epsilon$ for all maturities. 
\label{digital put}
\end{Proposition}

It seems intuitively plausible, and hence it is tempting to conjecture, that every economically admissible pricing kernel 
should have the UI property. For example, if the pricing kernel is a type-$D$ potential (Hunt \& Kennedy 2004, Rogers 1997, Rutkowski 1997), then it is UI (Meyer 1966).   
We proceed by establishing  the following for the Vasicek model. 
\begin{Proposition}
If $R_{\infty} > 0$ then $\{\pi_t\}$ is uniformly integrable. \label{Prop_2}
\end{Proposition}
\noindent {\em Proof}. 
We shall use an $\mathcal {L}^{\,p}$
test.  We say that a collection $\mathcal {C}$ of random variables is bounded in $\mathcal {L}^{\,p}$ if there exists a constant $\gamma >0$ such that ${\mathbb E}[|X|^p] < \gamma$ for all $X \in \mathcal {C}$. Now, if $p > 1$ and $x \geq \delta > 0$ for $x, \delta \in \mathbb R$, 
then clearly $x \leq \delta^{1-p} x^p $. It follows that if $\mathcal {C}$ is bounded in $\mathcal {L}^{\,p}$ then for all $X \in \mathcal {C}$ it holds that 
\begin{eqnarray}
{\mathbb E}[|X| {\mathds 1}\{|X|>\delta\}] \leq  \delta^{1-p} {\mathbb E}[|X^p| {\mathds 1}\{|X|>\delta\}] 
< \gamma  \delta^{1-p}.    
\end{eqnarray}
Thus given any $\epsilon > 0$ if we set 
\begin{eqnarray}
\delta = \left( \frac {\epsilon}{\gamma} \right) ^{1/(1-p)}
\end{eqnarray}
then we have constructed a $\delta$ such that (\ref{UI property_disp}) holds for all $X \in \mathcal {C}$.
Therefore, if a collection of random variables is bounded in $\mathcal {L}^{\,p}$ for some $p > 1$, then it is UI.  
It follows that a sufficient condition for the pricing kernel to be UI is that there should exist a $p>1$ and a $\gamma > 0$ such that ${\mathbb E}[\pi_t^p] < \gamma$
for all $t$. Now, a calculation starting with  (\ref{explicit pricing kernel}) gives
\begin{eqnarray}
\log {\mathbb E}[\pi_t^p] &=& -p\left[ \theta+\frac{1}{2}\lambda^2 - p \, \frac{1}{2}\left(\frac{\sigma}{k}-\lambda
\right)^2\right] t \nonumber \\ && + \frac{p}{k}\left[ \theta-r_0-p\frac{\sigma}{k} \left(\frac{\sigma}{k}-\lambda
\right)\right] \left( 1-\re^{-kt}\right)  + \frac{p^2\sigma^2}{4k^3} \left( 1-\re^{-2kt}\right) . 
\label{expectation in Lp test} 
\end{eqnarray}
The second and third terms on the right are bounded, so our goal is to show that if 
$R_{\infty} > 0$ then there exists a value of $p > 1$ such that the coefficient of $t$ in the first term on the right in
(\ref{expectation in Lp test}) is less than or equal to zero. Suppose therefore that $R_{\infty} > 0$. Then we have 
\begin{align}
   \theta > \frac{1}{2} \frac{\sigma^2}{k^2} - \frac{\lambda \sigma}{k}.
\end{align}
Completing the square on the right, we get 
\begin{align}
   \theta + \frac{1}{2}\lambda^2 > \frac{1}{2} \left( \frac{\sigma}{k} - \lambda \right)^2 .
\end{align}
Therefore if we set
\begin{eqnarray}
p = \frac{\theta+\frac{1}{2}\lambda^2}{ \frac{1}{2} \left(\frac{\sigma}{k}-\lambda\right)^2} , 
\label{value of p} 
\end{eqnarray}
then it follows that $p>1$ and that the first term on the right side of (\ref{expectation in Lp test}) vanishes. Since the other two terms are bounded, this shows that if $R_\infty>0$ then the $\mathcal {L}^{\,p}$ test is satisfied, and the pricing kernel is UI.  \hfill $\Box$ 

\begin{Proposition} \label{Prop_L1_BM}
If $R_{\infty} < 0$ then $\{\pi_t\}$ is not uniformly integrable.
\end{Proposition}
\noindent {\em Proof}. 
It is well known (Williams 1991) that if a collection of random variables is UI then it is bounded in $\mathcal {L}^{\,1}$. For suppose that $ \mathcal {C}$ has the property that for every $\epsilon>0$ there exists a $\delta \geq 0$ such that (\ref{UI property_disp}) holds for all $X \in \mathcal {C}$. Then there exists a constant $\delta_1$ such that ${\mathbb E}[|X| {\mathds 1}\{|X|>\delta_1\}] < 1$ for all $X \in \mathcal {C}$,
and therefore
\begin{eqnarray}
{\mathbb E}[|X| ] =
{\mathbb E}[|X| {\mathds 1}\{|X|>\delta_1\}] + {\mathbb E}[|X| {\mathds 1}\{|X| \leq \delta_1\}]
< 1 + \delta_1
\end{eqnarray} 
all $X \in \mathcal {C}$, and it follows that $\mathcal {C}$ is bounded in $\mathcal {L}^{\,1}$.
Thus to establish the statement of the proposition it suffices to show that if $R_{\infty} < 0$ then $\{ \pi_t \}$ is not bounded in $\mathcal {L}^{\,1}$. Keeping in mind that $\mathbb{E}[\pi_t] = P_{0t}$, we shall show that if $R_\infty < 0$, then for any choice of $\gamma >0$ there exists a time $t^*$ such that $P_{0t} > \gamma$ for all $t \geq t^*$. By virtue of  \eqref{expectation in Lp test}, 
we have
\begin{align}
P_{0t} = \exp \left[  -R_{\infty} t + \frac{1}{k} \left( 1 - {\rm e}^{-kt} \right) (R_{\infty} - r_0) - \frac{1}{4} \frac{\sigma^2}{k^3}\left( 1- {\rm e}^{-kt}\right)^2 \right]  . 
\label{discount bond price}
\end{align}
It follows that 
\begin{align}
\log P_{0t} \geq  -R_{\infty} t  + \frac{1}{k} (R_\infty - r_0)  \mathbbm{1} \{ R_{\infty} - r_0 \leq 0\}  - \frac{1}{4} \frac{\sigma^2}{k^3}.
\end{align} 
Therefore, let us define $t^*$ by setting
\begin{equation}
   -R_{\infty} t^*  + \frac{1}{k} (R_\infty - r_0)  \mathbbm{1} \{ R_{\infty} - r_0 \leq 0\}  - \frac{1}{4} \frac{\sigma^2}{k^3} = \log \gamma, 
\end{equation}
or equivalently
\begin{equation}
  t^* = \frac{1}{R_\infty} \left(  \frac{1}{k} (R_\infty - r_0)  \mathbbm{1} \{ R_{\infty} - r_0 \leq 0\}  - \frac{1}{4} \frac{\sigma^2}{k^3} - \log \gamma\right).
\end{equation}
Then it follows as a consequence that if $R_{\infty} <0$ then for all $t > t^*$ we have $P_{0t} > \gamma$. This shows that if $R_{\infty} < 0$ then the pricing kernel is not UI.  \hfill $\Box$ 

\begin{Proposition} 
If $R_{\infty} = 0$ then $\{\pi_t\}$ is not uniformly integrable.
\end{Proposition}
\noindent {\em Proof}. 
The situation when $R_{\infty} = 0$ is more delicate. The pricing kernel fails the 
$\mathcal {L}^{\,p}$ test if  $R_{\infty} = 0$, so we cannot conclude that it is UI. On the other hand, the pricing kernel is bounded in $\mathcal{L}^1$ if  $R_{\infty} = 0$, so we cannot conclude that it is \emph{not} UI. Thus when $R_{\infty} = 0$ the simple tests give us no information and we need to look at the definition of uniform integrability and ask whether \eqref{PK_UI} holds. We shall demonstrate that if $R_{\infty} = 0$ then \eqref{PK_UI} does not hold, and therefore the pricing kernel is not UI. First, let us define
\begin{eqnarray}
	\alpha_{st} = \frac{\sigma}{k} - \lambda - \frac{\sigma}{k} \,  \re^{k(s-t)}.
\end{eqnarray}
Using \eqref{PK_B_case} and \eqref{DB_BM} we can write
\begin{eqnarray}
	\pi_t = P_{0t} \exp\left( \int_0^t \alpha_{st} \, \rd W_s - \frac{1}{2} \int_0^t \alpha_{st}^2 \, \rd s\right).
\end{eqnarray}
Thus, for each value of $t$ the pricing kernel is of the form
\begin{eqnarray}
	\pi_t  = P_{0t} \exp\left( A_t Z - \frac{1}{2} A_t^2\right),
\end{eqnarray}
where $Z$ is normally distributed with mean zero and variance unity, and where we define $A_t$ (which we take to be positive) by 
\begin{eqnarray}
	A_t^2 = \int_0^t \alpha_{st}^2 \, \rd s.
\end{eqnarray}
It follows that
\begin{eqnarray}
	\mathbb{E}[\pi_t \mathds{1} \{ \pi_t > \delta \} ] = P_{0t} \, \mathbb{E}\left[ \exp\left( A_t Z - \frac{1}{2} A_t^2 \right) \mathds{1} \left\{ Z > \frac{\log \delta - \log P_{0t} + \half A_t^2}{A_t} \right\} \right].
\end{eqnarray}
The expectation can be computed by standard techniques, leading to the  following formula:
\begin{eqnarray}
\mathbb{E}[\pi_t \mathds{1} \{ \pi_t > \delta \} ] = P_{0t} \, N \left( \frac{ \log P_{0t} + \half A_t^2 - \log\delta}{A_t} \right).
\end{eqnarray}
Recall from \eqref{discount bond price} with $R_\infty = 0$ that 
\begin{equation}
  \log P_{0t} = -r_0 \frac{1}{k} \left(1 - {\rm e}^{-kt}\right)- \frac{1}{4} \frac{\sigma^2}{k^3} \left(1 - {\rm e}^{-kt}\right)^2, 
\end{equation}
which is bounded. We thus have
\begin{align}
  \sup_t \mathbb{E}[\pi_t \mathbbm{1}(\pi_t > \delta)] &\geq \exp \left[ \inf_u \log P_{0u} \right] \sup_t N \left( \frac{ \log P_{0t} + \half A_t^2 - \log\delta}{A_t} \right)  \nonumber \\
&\geq \exp \left[ \inf_u \log P_{0u} \right]  \sup_t N \left[\frac{ \inf_u ( \log P_{0u} )+ \half A_t^2 - \log \delta}{A_t}\right]. 
\label{inequality}
\end{align}
 It follows from
\begin{equation}
  \inf_t  \log P_{0t}= -\frac{r_0}{k} \mathbbm{1}(r_0 >0) - \frac{1}{4} \frac{\sigma^2}{k^3}
\end{equation}
that 
\begin{eqnarray}
\sup_t \mathbb{E}[\pi_t \mathbbm{1}(\pi_t > \delta)] &\geq&  \exp \left[ -\frac{r_0}{k} \mathbbm{1}(r_0 >0) - \frac{1}{4} \frac{\sigma^2}{k^3} \right] \nonumber \\  &\quad& \times \sup_t N \left[ \frac{1}{A_t}\left( -\frac{r_0}{k} \mathbbm{1}(r_0 >0) - \frac{1}{4} \frac{\sigma^2}{k^3} + \half A_t^2 - \log \delta \right)\right]. 
\end{eqnarray}
Since $\lim_{t \to \infty} A_t = \infty$, the supremum on the right  side is achieved in the limit as $t$ approaches infinity. As a consequence, we have
\begin{align}
 \sup_t \mathbb{E}[\pi_t \mathbbm{1}(\pi_t   > \delta)] 
 \geq \exp \left[ -\frac{r_0}{k} \mathbbm{1}(r_0>0) - \frac{1}{4} \frac{\sigma^2}{k^3}\right], 
\end{align}
which implies that
\begin{align}
  \lim_{\delta \to \infty} \sup_t \mathbb{E}[\pi_t \mathbbm{1}(\pi_t > \delta)] > 0,
\end{align}
and hence that the pricing kernel is not UI.
\hfill $\Box$ 

\section{Long-bond return process}
\noindent The return at time $t$ on an investment of one unit of account in the long bond is defined by the expression
\begin{equation}
	L_t = \lim_{T \to \infty} \frac{P_{tT}}{P_{0T}},
\end{equation}
providing the limit exists in a suitable sense (Flesaker \& Hughston 1996). We refer to $\{L_t\}_{t \geq 0}$ as the long-bond return process.
In the following, we consider the long-bond return process in the Vasicek model. We shall show that the relevant 
limit exists and that it can be worked out explicitly. Using the formula for the discount bond price, we find that
\begin{eqnarray}
\log \frac{P_{tT}}{P_{0T}} &=& R_\infty t + \frac{1}{k}(r_0 - r_t) - \frac{1}{k}(R_\infty - r_t) \re^{-k(T-t)} + 
\frac{1}{k} (R_\infty - r_0) \re^{-kT} \nonumber \\ && 
+ \frac{\sigma^2}{4 k^3}\left( \left(1- \re^{-kT}\right)^2 - \left(1 - \re^{-k(T-t)}\right)^2\right). 
\end{eqnarray}
One sees that the limit of this expression for large $T$ is given by 
\begin{eqnarray}
\lim_{T \to \infty} \log \frac{P_{tT}}{P_{0T}}  = R_\infty t +  \frac{1}{k}(r_0 - r_t).
\end{eqnarray}
It follows that the long-bond return process is
\begin{eqnarray}
L_t  = \exp \left[ {R_\infty t + \frac{1}{k}(r_0 - r_t)} \right]. 
\label{limit_proc}
\end{eqnarray}
If we recall expression (\ref{Vasicek pricing kernel with short rate}) for the pricing kernel in the Vasicek model, and expression 
(\ref{Vasicek long rate}) for the asymptotic rate, we deduce that the product of the pricing kernel and the long-bond return is given by  
\begin{eqnarray}
\pi_t L_t = \exp \left[ \left(\frac{\sigma}{k} - \lambda \right) W_t - \frac{1}{2} \left(\frac{\sigma}{k} - 
\lambda\right)^2 t \right]. 
\end{eqnarray}
This shows that the return on a unit investment in the long bond takes the form of a geometric Brownian motion asset in the Vasicek model, with volatility $\sigma/k$ and with a Vasicek-type 
integrated interest rate. More specifically, we have
\begin{eqnarray}
L_t =   \exp \left[ {\int_0^t \left(r_s + \frac{\lambda \sigma}{k}  \right)\rd s}  + {\frac{\sigma}{k}\, W_t - \frac{1}{2}
\left(\frac{\sigma}{k}\right)^2 t} \right].
\end{eqnarray}

The significance of the martingale $\{M_t\}_{t \geq 0}$ defined by $M_t =    \pi_t L_t $ is that it acts as the 
 change-of-measure density from the physical measure $\mathbb P$ to the so-called terminal measure (or long forward measure) introduced in Flesaker \& Hughston (1996). To see this, recall 
that to change from $\mathbb P$ to the measure associated with a given numeraire $\{N_t\}$, 
the change-of-measure martingale is given by $\{\pi_t N_t\}$. For example, to change from $\mathbb P$  to the risk-neutral measure associated with the use of the money market account as numeraire, 
the change-of-measure martingale is $\{\pi_t B_t\}$. In the present context,  the terminal measure 
agrees with $\mathbb P$  in the Vasicek model if and only if $M_t  = 1$ for $t \geq 0$, which holds if and only if 
\begin{equation}
\lambda = \frac{\sigma}{k} \,.
\end{equation}
The condition that the terminal measure and the physical measure agree has been shown in Qin \& Linetsky (2014) to be equivalent to the assumptions made in the recovery theorem of Ross (2015). Thus, we see that in the Vasicek model under the Ross recovery assumption one can 
infer the market price of risk from the current price levels of options on discount bonds, since such option prices depend on the ratio 
$\sigma/k$. However, there is no \emph{a priori} reason to believe that the interest-rate market price of risk should 
be equal in general to the volatility of the long-bond return process. In fact, we have the following: 
\begin{Proposition} 
For any arbitrage-free  interest-rate model based on a Brownian filtration, Ross recovery holds if and only if the interest-rate market price of risk is equal to the volatility of the long-bond return process. 
\end{Proposition}
\noindent {\em Proof}. We know that Ross recovery holds if and only if the terminal measure coincides with the physical measure, or equivalently $M_t = 1$ for all $t \geq 0$, which holds if and only if $L_t = 1/ \pi_t$ for all $t \geq 0$, which holds if and only if the the interest-rate market price of risk agrees with the volatility of the return on  the long bond.
\hfill $\Box$ 

\vspace{3mm}

\noindent This observation suggests that it is unlikely that Ross recovery will be observed in financial markets, a view supported by empirical evidence (Borovi\v{c}ka \emph{et al} 2014, Qin \emph{et al} 2016). 

\section{Geometric L\'evy models}

\noindent
Are the foregoing conclusions---in particular, those addressing the feasibility of Ross recovery and those relating the positivity of the long rate of interest to the uniform integrability of the pricing kernel---specific to markets based on Brownian filtrations? To investigate this question, we consider the more general case of markets based on L\'evy filtrations. 

To begin, it will be useful if we briefly review the pricing kernel approach to geometric L\'evy models. In such models, the pricing kernel method has the advantage that it brings to light the relations between risk, risk aversion, and return for models with price jumps (Brody \textit{et al} 2012). 
Let $\{\xi_t\}$ be a L\'evy process, and $\lambda>0$ the level of risk aversion. We assume that  $\{\xi_t\}$ satisfies a moment condition of the form
\begin{equation}{\mathbb E}\left[\exp{\alpha \xi_t}\right] < \infty \quad \mbox{ for }  t \geq 0 \mbox{ and } \alpha\in A, \label{mom_cond} 
\end{equation}
 for some connected set $A \subset \mathbb {R}$ containing the origin. The pricing kernel 
 of a geometric L\'evy model, with constant interest rate $r$, is given by 
\begin{equation}
\pi_t = \re^{-rt} \, \re^{-\lambda \xi_t - t \psi(-\lambda)} .  
\end{equation}
Here $ \{ \psi(\alpha) \}_{ \alpha \in A }$,  is the so-called L\'evy exponent, defined by 
\begin{equation}
{\mathbb E}[\re^{\alpha \xi_t}] = 
\re^{\psi(\alpha) \,t}. 
\label{Levy exponent}
\end{equation}
It is straightforward to check that the L\'evy exponent is a convex function. Since the product of the pricing kernel and the price $\{S_t\}$ of a non-dividend-paying asset is a $\mathbb{P}$-martingale, we may suppose that there is a $\beta \in A$ such that
\begin{equation}
	\pi_t S_t = S_0 \re^{\beta \xi_t - t \psi(\beta)}.
\end{equation}
Writing $\sigma = \beta + \lambda$, we thus deduce that
\begin{equation}
S_t = S_0 \, \re^{ rt + R(\lambda,\sigma)t + \sigma \xi_t - t \psi(\sigma)} , 
\label{Levy asset price}
\end{equation}
where the excess rate of return $R(\lambda, \sigma)$ is given by
\begin{equation}
{R}(\lambda,\sigma) = \psi(\sigma) + \psi(-\lambda) - \psi(\sigma-\lambda).
\label{eq:35} 
\end{equation} 
A short calculation shows that $R(\lambda, \sigma)$ is bilinear in $\lambda$ and $\sigma$ if and only if $\{\xi_t\}$ 
is a Brownian motion (Brody \textit{et al} 2012). It follows that the widely popularized interpretation of $\lambda$ as a ``market price of risk'', which is valid for models based on a Brownian filtration, does not 
carry through to the general L\'evy regime. Nevertheless, the notion of excess rate of return is well defined, 
and  the strict convexity of the L\'evy exponent implies that the excess rate of return is both an 
increasing function of $\lambda$ and an increasing function of $\sigma$. 

It is worth drawing attention to the fact that the value of the asset given by \eqref{Levy asset price} does not depend on the drift of the L\'evy process.  Therefore without loss of generality we can set the drift of the L\'evy process equal to zero. In that case we refer to $\{\xi_t\}$ as a compensated L\'evy process.  This implies that  ${\mathbb E}[\xi_t] = 0$ and that $\{\xi_t\}$ is 
a martingale. For example, if $\{N_t\}$ is the standard Poisson process, with jump rate 
$\mu$, then the associated compensated L\'evy process  is given by $\xi_t = N_t - \mu t$.  With these conventions in mind we proceed to establish the following lemmas, which will be useful in what follows. 

\begin{Lemma} \label{psi_lemma}
	Let $\{\psi(\alpha)\}_{\alpha \in A}$ be the L\'evy exponent of a compensated L\'evy process $\{\xi_t\}$ that  admits exponential moments for some connected set $A \subset \mathbb {R}$ containing the origin. Then $\psi$ is strictly positive on its domain of definition, except at the origin, where it vanishes.
\end{Lemma}
\noindent {\em Proof.} Differentiating each side of \eqref{Levy exponent} and setting $\alpha = 0$, we obtain 
$\mathbb{E}[\xi_t] = \psi'(0) t$ for all $t \geq 0$. Since $\{\psi(\alpha)\}$ is by assumption a compensated L\'evy process, it follows that $\psi'(0) = 0$.  
Hence, the curve $\psi : A \rightarrow \mathbb R$ defined by $\alpha \in A \rightarrow  \psi(\alpha)$ has a horizontal tangent at the origin. 
Since $\psi$ is strictly convex, and thus lies above any of its tangents except at the point where the tangent touches the curve, we conclude that $\psi$ is strictly positive except at the origin. At the origin, we have $\psi(0) = 0$, which follows from the definition of the L\'evy exponent. \hfill $\Box$

\vspace{3mm}

\begin{Lemma} \label{nonneg}
	Let $A \subset \mathbb{R}$ be a connected set containing the origin, and let $f: A \to \mathbb{R}$ be a nonnegative strictly convex function, differentiable on $A$ and vanishing at $0$. Then it holds that $xf'(x)  > f(x)$ for all $x \in A$ except at $x = 0$. 
	\end{Lemma}
\noindent {\em Proof.} Let $x \in A$. If $ x > 0$, then by the mean value theorem there exists $y \in (0, x)$ such that $f(x) = x f'(y)$. Since $f$ takes its minimum at the origin and is strictly convex, it follows that $f'(y) < f(x)$. Therefore, $f(x) < x f'(x)$, as required. On the other hand, if $ x < 0$, then the mean value theorem says that there exists $y \in (x, 0)$ such that $f(x) = x f'(y)$. But since $f$ is strictly convex with a minimum at the origin, it follows that $0 < f'(x) < f'(y)$, and thus $x f'(y) < x f'(x)$, since $x < 0$. Therefore, $f(x) < x f'(x)$. \hfill $\Box$

\vspace{3mm}

\section{Construction of L\'evy-Vasicek model}

\noindent
Our objective going forward is to investigate properties of the pricing kernel in the L\'evy analogue of the Vasicek model. 
Specifically, we are interested in an Ornstein-Uhlenbeck-type short rate model driven by a L\'evy process. 
Such models have been investigated, for example, in Norberg (2004). Remarkably, the condition $\lambda = \sigma/k$ for Ross recovery in the Vasicek model is unchanged in its L\'evy-Vasicek 
counterpart. We shall also show that the pricing kernel is UI if and only if the long rate of interest is strictly positive. 

The pricing kernel method allows us to work out the details of the general L\'evy-Vasicek model 
in the  ${\mathbb P}$-measure. We fix a probability space $(\Omega, \mathcal{F}, \mathbb P)$ and introduce a one-dimensional L\'evy process $\{\xi_t\}_{t \geq 0}$. We assume that \eqref{mom_cond} holds for some connected set $A \subset \mathbb {R}$ containing the origin, and we write $\psi (\alpha)$ for the L\'evy exponent, defined for $\alpha\in A$. 
 The pricing kernel in the L\'evy-Vasicek model then takes the form
\begin{equation}
\pi_t = \exp \left[ -\int_0^t r_s \, {\rm d}s -\lambda \xi_t - \psi(-\lambda) \, t \right], 
\label{Levy pricing kernel}
\end{equation} 
where the short rate is a L\'evy-Ornstein-Uhlenbeck process satisfying a dynamical equation of the form
\begin{equation}
\rd r_t = k(\theta - r_t) \, \rd t - \sigma \rd \xi_t. 
\end{equation}

The parameters of the model here have essentially the same interpretation as those of the classical Vasicek model. The only difference is that in the L\'evy case we let the volatility parameter have dimensions of inverse time. The L\'evy process itself is taken to be dimensionless. This is of course quite natural in the case of counting processes. In the situation where $\{\xi_t\}$ is a Brownian motion, it will be understood that a standard Brownian motion is multiplied by an appropriate volatility parameter to make the overall process dimensionless.  Without loss of generality we can set the drift of the L\'evy process equal to zero by absorbing any drift into 
the definition of the mean-reversion level.  Thus in what follows we assume that $\{\xi_t\}$ is a compensated L\'evy process. As in the Brownian case, we find by use of an integrating factor that the short rate  is given by 
\begin{equation}
r_t = \theta + (r_0 - \theta) \re^{-kt} - \sigma \int_0^t \re^{k(s-t)} \rd \xi_s .
\label{Levy short rate}
\end{equation}
The integrated short rate can be worked out by a calculation that parallels that of the classical Vasicek model, with the following result:
\begin{equation}
\int_0^t r_s \rd s = \theta t + \frac{1}{k}(r_0 - r_t) - \frac{\sigma}{k} \xi_t \,. 
\label{Rt_Levy}
\end{equation}
It follows that the pricing kernel in the L\'evy-Vasicek model can be written in the form
\begin{equation}
\pi_t =\exp  \left [-\big(\theta + \psi(-\lambda) \big)t + \left(\frac{\sigma}{k} - \lambda\right)\xi_t - 
\frac{1}{k}(r_0 - r_t) \right ].
\end{equation}
Alternatively, if we insert the expression for $r_t$ given in \eqref{Levy short rate} then we obtain
\begin{equation}
 \pi_t = \exp  \left[  -\left(\theta +  \psi(-\lambda) \right) t - \frac{1}{k} \left(1 - {\rm e}^{-kt}\right)(r_0 - \theta) + \int_0^t \left( \frac{\sigma}{k} - \lambda - \frac{\sigma}{k} {\rm e}^{k(u-t)} \right) \rd \xi_u \right].
 \label{Levy-Vasicek pricing kernel}
\end{equation}
\section{Discount bonds in L\'evy-Vasicek model}
\noindent
We proceed to obtain an expression for the price of a discount bond. By  \eqref{Levy-Vasicek pricing kernel} we have
\begin{equation}
  \log \pi_T = -\left(\theta + \psi(-\lambda)\right) t - \frac{1}{k} \left(1 - {\rm e}^{-kT}\right)(r_0 - \theta) + \int_0^T \left( \frac{\sigma}{k} - \lambda - \frac{\sigma}{k} {\rm e}^{k(u-T)} \right) \rd \xi_u,
\end{equation}
from which we deduce that
\begin{align}
  \log \pi_T &= -\left(\theta + \psi(-\lambda) \right) T- \frac{1}{k} \left( 1- {\rm e}^{-kT}\right) (r_0 - \theta) + 
\int_0^t \left( \frac{\sigma}{k} - \lambda - \frac{\sigma}{k} {\rm e}^{k(u-T)} \right) \rd \xi_u  \nonumber \\
&\quad \quad + \int_t^T \left( \frac{\sigma}{k} - \lambda - \frac{\sigma}{k} {\rm e}^{k(u-T)} \right) \rd \xi_u.
\end{align}
Hence, 
\begin{align}
  \mathbb{E}_t [\pi_T] &= \exp \left[  -\left(\theta +\psi(-\lambda)\right) T - \frac{1}{k} \left( 1- {\rm e}^{-kT}\right) (r_0 - \theta) + 
\int_0^t \left( \frac{\sigma}{k} - \lambda - \frac{\sigma}{k} {\rm e}^{k(u-T)} \right) \rd \xi_u \right]  \nonumber \\
&\quad \quad \times \mathbb{E}_t \exp\left[ \int_t^T \left( \frac{\sigma}{k} - \lambda - \frac{\sigma}{k} {\rm e}^{k(u-T)} \right) \rd \xi_u \right].
\end{align}
To work out the conditional expectation on the right side we use the identity 
\begin{equation}
\mathbb{E}_t \exp \left[{\int_t^T\alpha_s {\rm d} \xi_s}\right] =  \exp {\int_t^T \psi(\alpha_s) {\rm d} s} , 
\end{equation}
valid for any Borel  function $\{\alpha_s\}_{s\geq 0}$ taking values in the interval $A$. 
It follows that
\begin{align}
  \mathbb{E}_t \exp\left[ \int_t^T \left( \frac{\sigma}{k} - \lambda - \frac{\sigma}{k} {\rm e}^{k(u-T)} \right) \rd \xi_u \right] = \exp \left[  \int_t^T \psi \left( \frac{\sigma}{k} - \lambda - \frac{\sigma}{k} {\rm e}^{k(u-T)} \right) \rd u\right].
\end{align}
Therefore, by expression (\ref{discount bond formula}) for the discount bond we obtain
\begin{align}
  \log P_{tT} &=  -\left( \theta + \psi (- \lambda) \right) (T - t) - \frac{1}{k} \left( {\rm e}^{-kt} - {\rm e}^{-kT}\right)(r_0 - \theta) \nonumber \\
&\quad + \int_t^T \psi \left( \frac{\sigma}{k} - \lambda - \frac{\sigma}{k} {\rm e}^{k(u-T)} \right) \rd u + \frac{\sigma}{k} \left(1 - {\rm e}^{k(t-T)}\right) \int_0^t {\rm e}^{k(u-t)} \rd \xi_u.
\label{canceling terms}
\end{align}
As a consequence of \eqref{Levy short rate} we can write
\begin{equation}
  \sigma \int_0^t {\rm e}^{k(u-t)} \rd \xi_u = \theta + (r_0 - \theta) {\rm e}^{-kt} - r_t.
\end{equation}
Thus, the terms involving $r_0 - \theta$ in \eqref{canceling terms} cancel, and we are left with the following expression for the price of a $T$-maturity discount bond: 
\begin{align}
P_{tT} = \exp \left [ -\left( \theta + \psi (\lambda)\right) (T - t) 
 +  \int_t^T \psi(\alpha_{uT}) \, \rd u + \frac{1}{k} \left(1 - {\rm e}^{k(t-T)}\right) \left(\theta  - r_t\right) \right], 
 \label{log Levy bond price}
\end{align}
where for $0 \leq s \leq t$ we set
\begin{equation}
\alpha_{ut} =  \frac{\sigma}{k} - \lambda -\frac{\sigma}{k}\, \re^{k(u- t)} . 
\label{Levy alpha} 
\end{equation}
To investigate the asymptotic bond yield, or exponential long rate, first we show that
  \begin{equation}
  \lim_{T \to \infty} \frac{1}{T-t}  \int_t^T \psi\left(  \alpha_{sT} \right) \rd s = \psi\left( \frac{\sigma}{k} - \lambda\right). \label{limit}
\end{equation}
We note that the derivative  of the numerator with respect to $T$ is given by
\begin{eqnarray}
  \frac{\rm d}{\rd T}  \int_t^T \psi(  \alpha_{sT} ) \, \rd s &=&  \psi(\alpha_{TT}) + \sigma \int_t^T \psi'(\alpha_{sT}) {\rm e}^{k(s-T)} \rd s \nonumber \\
&=&  \psi(\alpha_{TT}) - \int_t^T \frac{\rm d}{\rd s} \psi(\alpha_{sT})  \, \rd s \nonumber \\
&=& \psi(\alpha_{tT}).
\end{eqnarray}
Thus, applying l'Hospital's rule we obtain
\begin{equation}
  \lim_{T \to \infty} \frac{1}{T-t}  \int_t^T \psi\left(  \alpha_{sT} \right) \rd s = \lim_{T \to \infty} \psi(\alpha_{tT}) = \psi\left( \frac{\sigma}{k} - \lambda\right),
\end{equation}
establishing \eqref{limit}.
One sees that in the L\'evy-Vasicek model, as in the Brownian case, the long rate does not depend on $t$, and we have the following: 
\begin{eqnarray}
R_{\infty} = - \lim_{T \to \infty} \frac{1}{T-t} \log P_{tT} = \theta + \psi(-\lambda) - \psi\left(\frac{\sigma}{k} - \lambda\right).
\end{eqnarray}

\section{Uniform integrability in L\'evy-Vasicek model}
\noindent
A natural question that emerges, in view of our conjecture that an economically admissible interest rate model should be 
accompanied by a UI pricing kernel, is whether and for what choice of parameters the pricing kernel in a  
L\'evy-Vasicek model is UI. For this purpose, it will be useful to express the pricing kernel in the form 
\begin{eqnarray}
\pi_t &=& \exp\left( -\big( \theta + \psi(-\lambda)\big)t -\frac{1}{k} (r_0 -\theta)(1- \re^{-kt}) + 
\int_0^t \alpha_{st} \,  \rd \xi_s \right),  
\label{Levy_PK_UI1}
\end{eqnarray} 
or equivalently
\begin{eqnarray}
\pi_t 
&=&  \exp\left(-R_\infty t - \psi\left(\frac{\sigma}{k} - \lambda\right)t -  \frac{1}{k} (r_0 -\theta)(1- \re^{-kt}) + 
\int_0^t   \alpha_{st} \, \rd \xi_s\right), \label{Levy_PK_UI}
\end{eqnarray} 
where we recall the definition \eqref{Levy alpha} for $\alpha_{st}$. It follows that 
\begin{eqnarray}
\log {\mathbb E}[\pi_t] &=& -R_\infty t -  \frac{1}{k} (r_0 -\theta)(1- \re^{-kt}) + 
\int_0^t \psi\left(  \alpha_{st} \right) \rd s  - \psi\left(\frac{\sigma}{k} - \lambda\right) t. \label{logex}
\end{eqnarray} 

In the asymptotic considerations that follow, we adopt the following conventions. Given a pair of functions
$f: \mathbb R^+ \to \mathbb R$  and $g: \mathbb R^+ \to \mathbb{R}\backslash \{0\}$, we say that $f$ is $O(g)$ for large $t$ if
\begin{eqnarray}
  \limsup_{t \to \infty}  \left|  \frac{f(t)}{g(t)} \right|< \infty,
\end{eqnarray}
and we say that $f$  is $o(g)$ for large $t$ if
\begin{eqnarray}
  \lim_{t \to \infty}  \frac{f(t)}{g(t)}  =0.
\end{eqnarray}
With reference to the integral appearing on the right side of \eqref{logex}, we shall show that
\begin{equation}
  \int_0^t \psi\left(  \alpha_{st} \right) \rd s = \psi \left( \frac{\sigma}{k} - \lambda\right)t + O(1) 
\label{Lemma2_eq}
\end{equation}
for large $t$.  We note that
\begin{eqnarray}
  \left|   \int_0^t \left( \psi\left(  \alpha_{st} \right) - \psi\left(\frac{\sigma}{k} - \lambda\right) \right)\rd s \right| \leq    \int_0^t \left| \psi\left(  \alpha_{st} \right) - \psi\left(\frac{\sigma}{k} - \lambda\right) \right|\rd s. \label{estimate1}
\end{eqnarray}
But it follows from the mean value theorem that, for a fixed value of $s$,  there exists a $\rho$ in the open interval 
$( -  \sigma k^{-1} {\rm e}^{k(s - t)}, 0)$
such that
 \begin{eqnarray}
   \psi \left( \frac{\sigma}{k} - \lambda\right)
 = \psi(\alpha_{st}) + \frac{\sigma}{k} \, {\rm e}^{k(s-t)} \psi'\left( \frac{\sigma}{k} - \lambda + \rho\right).
 \end{eqnarray}
Hence,
\begin{eqnarray}
   \left| \psi\left(  \alpha_{st} \right) - \psi\left(\frac{\sigma}{k} - \lambda\right) \right| = \frac{\sigma}{k} \, {\rm e}^{k(s-t)} \left|  \psi'\left( \frac{\sigma}{k} - \lambda + \rho\right) \right|.
\end{eqnarray}
Recall from Lemma \ref{psi_lemma} that $\psi$ is a nonnegative strictly convex function taking its minimum value at zero. Thus, over the relevant range of $\rho$, the maximum of $|\psi'(\sigma/k - \lambda + \rho)|$ is taken, depending on the value of $s$, either at the left boundary $\rho = -  (\sigma / k) {\rm e}^{k(s - t)}$ or at the right boundary $\rho = 0$. More precisely, there exists a $t' \in [0, t]$ such that as $s$ varies, the maximum is achieved at the right boundary whenever $s \in (0, t')$, and at the left boundary whenever $s \in (t', t)$. Thus, when $s \in (0, t')$ we have
\begin{eqnarray}
   \left| \psi\left(  \alpha_{st} \right) - \psi\left(\frac{\sigma}{k} - \lambda\right) \right| \leq \frac{\sigma}{k} {\rm e}^{k(s-t)} \left| \psi'\left( \frac{\sigma}{k} - \lambda\right)\right|,
\end{eqnarray}
and when $s \in (t', t)$ we have
 \begin{eqnarray}
   \left| \psi\left(  \alpha_{st} \right) - \psi\left(\frac{\sigma}{k} - \lambda\right) \right| \leq \frac{\sigma}{k} {\rm e}^{k(s-t)}  \left|\psi'\left( \frac{\sigma}{k} - \lambda - \frac{\sigma}{k} {\rm e}^{k(s-t)} \right)\right| \leq \frac{\sigma}{k} {\rm e}^{k(s-t)}  \left|\psi'(-\lambda)\right|.
\end{eqnarray}
In the last step here we have made use of the fact that since $\psi$ is convex, $|\psi'|$ is decreasing on the negative half line.
Reverting to the right side of \eqref{estimate1}, we thus deduce that
\begin{eqnarray}
   \int\limits_0^t \left| \psi\left(  \alpha_{st} \right) - \psi\left(\frac{\sigma}{k} - \lambda\right) \right|\rd s &=& 
   \int\limits_0^{t'} \left| \psi\left(  \alpha_{st} \right) - \psi\left(\frac{\sigma}{k} - \lambda\right) \right|\rd s +  
   \int\limits_{t'}^{t} \left| \psi\left(  \alpha_{st} \right) - \psi\left(\frac{\sigma}{k} - \lambda\right) \right|\rd s \notag \\ \nonumber 
      &\leq& \frac{\sigma}{k}   \psi'\left( \frac{\sigma}{k} - \lambda\right) \int_0^{t'}   {\rm e}^{k(s-t)} \rd s  + \frac{\sigma}{k}  \left|\psi'\left( - \lambda  \right)\right|   \int_{t'}^t   {\rm e}^{k(s-t)}  \rd s\\
&=& \frac{\sigma}{k^2}  \psi'\left( \frac{\sigma}{k} - \lambda\right) \re^{-kt} \left(\re^{kt'} - 1\right) + \frac{\sigma}{k^2} |\psi'(-\lambda)| \left(1 - \re^{-k(t-t')}\right) \notag \\
&\leq& \frac{\sigma}{k^2}  \psi'\left( \frac{\sigma}{k} - \lambda\right) +  \frac{\sigma}{k^2} |\psi'(-\lambda)|,
\end{eqnarray}
and hence
\begin{equation}
    \left|   \int_0^t  \psi ( \alpha_{st} ) \, \rd s - \psi\left(\frac{\sigma}{k} - \lambda\right) t \right| \leq \frac{\sigma}{k^2}  \psi'\left( \frac{\sigma}{k} - \lambda\right) +  \frac{\sigma}{k^2} |\psi'(-\lambda)|.
\end{equation}
It  follows that
\begin{equation}
	\limsup_{t \to \infty}   \left|   \int_0^t  \psi ( \alpha_{st} ) \, \rd s - \psi\left(\frac{\sigma}{k} - \lambda\right) t \right| < \infty,
\end{equation}
which establishes \eqref{Lemma2_eq}.

With these preparations at hand, we are ready to return to our considerations of the uniform integrability of the pricing kernel.
\begin{Proposition}
If $R_{\infty} > 0$ then $\{\pi_t\}$ is uniformly integrable.
\end{Proposition}
\noindent {\em Proof.} We shall use an $\mathcal {L}^{\,p}$ test. Specifically, we show that if $R_{\infty} > 0$ then  there exists a $p > 1$ such that $\sup_t \mathbb{E}[\pi^p_t] < \infty$. 
We claim that it suffices to prove that if $R_\infty > 0$ then $\lim_{t \to \infty} \mathbb{E}[\pi^p_t] = 0$ for some $p > 1$. To see this, note that if $\lim_{t \to \infty} \mathbb{E}[\pi^p_t] = 0$ holds for some $p > 1$, there exist positive constants $C$ and $T$ such that $\mathbb{E}[\pi^p_t] < C$ for all $t \geq T$. But then
\begin{equation}
  \sup_t \mathbb{E}[\pi_t^p] \leq \max\left(C, \sup_{t \leq T} \mathbb{E}[\pi_t^p]\right).
\end{equation}
Since continuous functions are bounded on compact intervals, we see that $\sup_{t \leq T} \mathbb{E}[\pi_t^p]$ is bounded, and thus $ \sup_t \mathbb{E}[\pi_t^p] < \infty$, as required.
 Let us therefore show that $\lim_{t \to \infty} \mathbb{E}[\pi^p_t] = 0$ for some $p > 1$. It follows from \eqref{Levy_PK_UI} after a calculation that 
\begin{eqnarray}
\log {\mathbb E}[\pi_t^p] &=& -pR_\infty t -  \frac{p}{k} (r_0 -\theta)(1- \re^{-kt}) + 
\int_0^t \psi\left( p \alpha_{st} \right) \rd s  -p \psi\left(\frac{\sigma}{k} - \lambda\right) t,
\end{eqnarray} 
and we can use \eqref{Lemma2_eq} to see that
\begin{eqnarray} \label{general_eq}
  \log {\mathbb E}[\pi_t^p] &=&  -pR_\infty t -  \frac{p}{k} (r_0 -\theta)(1- \re^{-kt}) \nonumber \\
&\quad& \quad + 
\left(\psi\left( p\left( \frac{\sigma}{k} - \lambda \right)\right)  -p \psi\left(\frac{\sigma}{k} - \lambda\right)\right) t + O(1)
\end{eqnarray}
for large $t$. Since $\psi$ is continuous at $\sigma/k - \lambda$, we observe that the quantity
\begin{eqnarray}
  \left| \left(\psi\left( p\left( \frac{\sigma}{k} - \lambda \right)\right)  -p \psi\left(\frac{\sigma}{k} - \lambda\right)\right) \right|
\end{eqnarray}
can be made arbitrarily small as $p$ approaches unity from above. Keeping in mind that $R_\infty >0$, we thus see that there exists a $p > 1$ such that
\begin{eqnarray}
  -R_\infty + \left(\psi\left( p\left( \frac{\sigma}{k} - \lambda \right)\right)  -p \psi\left(\frac{\sigma}{k} - \lambda\right)\right) < 0,
\end{eqnarray}
and hence  $\lim_{t \to \infty} \mathbb{E}[\pi_t^p] = 0$, which is the required result. \hfill $\Box$

\begin{Proposition}
   If $R_{\infty} < 0$ then $\{\pi_t\}$ is not uniformly integrable.
\end{Proposition}
\noindent {\em Proof.} 
We shall show that $\{\pi_t\}$ is not bounded in $\mathcal {L}^{\,1}$, which implies that $\{\pi_t\}$ is not UI. It follows from \eqref{logex} and \eqref{Lemma2_eq} that for large $t$ one has
\begin{eqnarray} 
  \log {\mathbb E}[\pi_t] =  -R_\infty t -  \frac{1}{k} (r_0 -\theta)(1- \re^{-kt})  + O(1).
\end{eqnarray}
If $R_\infty < 0$, we have 
\begin{eqnarray} 
\lim_{t \to \infty} \mathbb{E}[\pi_t] = \infty.
\end{eqnarray}
Thus, the pricing kernel is not bounded in   $\mathcal {L}^{\,1}$. 
\hfill $\Box$
\begin{Proposition} 
If $R_{\infty} = 0$ then $\{\pi_t\}$ is not uniformly integrable. \label{Levy prop}
\end{Proposition}
\noindent {\em Proof}.  We recall that the pricing kernel is UI if and only if \eqref{PK_UI} holds.
Using \eqref{log Levy bond price} and \eqref{Levy_PK_UI1} we have
\begin{eqnarray}
	\pi_t = P_{0t} \exp\left( \int_0^t \alpha_{st} \,  \rd \xi_s - \int_0^t \psi(\alpha_{st})\,  \rd s\right).
\end{eqnarray}
As a consequence, we see that 
\begin{eqnarray}
	\mathbb{E}[\pi_t \mathds{1} \{ \pi_t > \delta \} ] =  P_{0t} \mathbb{E}\bigg[ \exp\left( \int_0^t \alpha_{st} \,  \rd \xi_s - \int_0^t \psi(\alpha_{st}) \, \rd s\right) \mathds{1} \left\{  \int_0^t \alpha_{st} \, \rd \xi_s > B(t, \delta) \right\} \bigg], \nonumber \\
\end{eqnarray}
where we define
\begin{eqnarray}
B(t, \delta) =  \log \delta - \log P_{0t} + \int_0^t \psi(\alpha_{st}) \, \rd s.
\end{eqnarray}
Equivalently, by \eqref{log Levy bond price} we have
\begin{eqnarray}
B(t, \delta) = \log \delta + R_{\infty} t + \frac{1}{k}(r_0 - \theta) \left(1 - {\rm e}^{-kt}\right) + \psi\left(\frac{\sigma}{k} - \lambda \right) t.
\label{barrier}
\end{eqnarray}
We introduce a new measure $\mathbb{P}^*$ on $\mathcal{F}_t$ by setting
\begin{eqnarray}
	\mathbb{P}^*(A) = \mathbb{E} \left[ \exp \left( \int_0^t \alpha_{st} \, \rd \xi_s -\int_0^t \psi(\alpha_{st})  \, \rd s \right) \mathds{1}\{A\} \right]
\end{eqnarray}
for $A \in \mathcal{F}_t$. Writing $\mathbb{E}^*$ for expectation under $\mathbb{P}^*$ we then have
\begin{eqnarray}
  \mathbb{E}[\pi_t \mathds{1} \{ \pi_t > \delta \} ] &=& P_{0t}  \, \mathbb{E}^{*}\left[  {\mathds 1}\left\{ \int_0^t \alpha_{st} \, \rd \xi_s > B(t, \delta) \right\}\right]. \label{exp_term}
\end{eqnarray}
Let us introduce a positive constant $\omega$ with units of inverse time, making $\omega  t$ dimensionless. Thus $\omega$ is a fixed ``rate". 
To proceed, we need the following results regarding the mean and variance of the random variable 
 \begin{eqnarray}
\int_0^t \alpha_{st} \rd \xi_s 
\end{eqnarray}
under $\mathbb{P}^*$. If $R_\infty = 0$, then for large $t$ we have
 \begin{eqnarray}
   \mathbb{E}^{*} \left[\int_0^t \alpha_{st} \, \rd \xi_s\right] - 	B(t, \omega t) =Ct + o(t), \label{help_lemma1_eq}
\end{eqnarray}
where 
\begin{eqnarray}
	C = \left(\frac{\sigma}{k} - \lambda\right) \psi'\left(\frac{\sigma}{k} - \lambda\right) - \psi\left(\frac{\sigma}{k} - \lambda\right) 
\label{help_lemma1_eq2}
\end{eqnarray}
is a positive constant, and
\begin{eqnarray}
  {\rm Var}^{*} \left[ \int_0^t \alpha_{st} \, \rd \xi_s\right] = O(t). \label{Var}
\end{eqnarray}
To see \eqref{help_lemma1_eq}, \eqref{help_lemma1_eq2}, and \eqref{Var}, note that for $u$ sufficiently close to zero we have
\begin{eqnarray}
  \mathbb{E}^{*}\left[ \exp\left( u \int_0^t \alpha_{st}\, \rd \xi_s\right)\right] = \exp \left( \int_0^t \left(\psi\left( (1 + u) \alpha_{st}\right) - \psi(\alpha_{st})\right)\rd s \right). \label{mom_gen}
\end{eqnarray}
Taking a derivative with respect to $u$ on both sides of this equation and setting $u = 0$ gives 
\begin{eqnarray}
  \mathbb{E}^{*} \left[\int_0^t \alpha_{st} \, \rd \xi_s\right] = \int_0^t \psi'(\alpha_{st}) \alpha_{st} \, \rd s.
\end{eqnarray}
A calculation then shows that
\begin{eqnarray}
   \frac{\rd}{\rd t}  \mathbb{E}^{*} \left[\int_0^t \alpha_{st} \, \rd \xi_s\right] &=& \alpha_{0t} \psi'(\alpha_{0t}), \label{expect_star}
\end{eqnarray}
and thus, by l'Hospital's rule,
\begin{eqnarray}
	\lim_{t\to \infty} \frac{1}{t} \mathbb{E}^{*} \left[\int_0^t \alpha_{st} \, \rd \xi_s\right] = \lim_{t\to \infty} \alpha_{0t} \,\psi'(\alpha_{0t}) = \left(\frac{\sigma}{k} - \lambda\right) \psi'\left(\frac{\sigma}{k} - \lambda\right).
\end{eqnarray}
It follows that 
\begin{eqnarray}
     \mathbb{E}^{*} \left[\int_0^t \alpha_{st} \, \rd \xi_s\right] &=&  \left(\frac{\sigma}{k} - \lambda\right) \psi'\left(\frac{\sigma}{k} - \lambda\right) t + o(t) \label{linear_exp}
\end{eqnarray}
for large $t$. Recall from \eqref{barrier} that $B(t, \delta)$ grows to leading order like $\psi(\sigma/k - \lambda) t$ when $R_\infty = 0$. This remains the case if we replace $\delta$ by $\omega t$, since the growth of $ \log \omega t$ is slower than linear growth. 
 That is, 
\begin{eqnarray}
   \mathbb{E}^{*} \left[\int_0^t \alpha_{st}\,  \rd \xi_s\right] - 	B(t, \omega t)  = \left[ \left(\frac{\sigma}{k} - \lambda\right) \psi'\left(\frac{\sigma}{k} - \lambda\right) - \psi\left(\frac{\sigma}{k} - \lambda\right)\right] t + o(t)
\end{eqnarray}
for large $t$. The positivity of the coefficient of $t$ on the right  side follows from Lemma \ref{nonneg}. We have thus arrived at equations \eqref{help_lemma1_eq} and \eqref{help_lemma1_eq2}. It remains to show equation \eqref{Var}. Taking the second  derivative of \eqref{mom_gen} with respect to $u$  and setting $u = 0$, we find
 \begin{eqnarray}
   \mathbb{E}^{*} \left[\left( \int_0^t \alpha_{st} \, \rd \xi_s\right)^2\right] = 
\left(\int_0^t \psi'(\alpha_{st}) \alpha_{st}\,  \rd s\right)^2 + \int_0^t \psi''(\alpha_{st}) \alpha_{st}^2 \, \rd s.
 \end{eqnarray}
Using \eqref{expect_star}, we then obtain
\begin{eqnarray}
    {\rm Var}^{*} \left[ \int_0^t \alpha_{st} \,  \rd \xi_s\right] &=& \mathbb{E}^{*} \left[\left( \int_0^t \alpha_{st} \, \rd \xi_s\right)^2\right] - \left( \mathbb{E}^{*} \left[ \int_0^t \alpha_{st} \, \rd \xi_s\right] \right)^2  \nonumber \\ &=&  \int_0^t \psi''(\alpha_{st}) \alpha_{st}^2  \, \rd s.
\end{eqnarray}
The limit
\begin{eqnarray}
\lim_{t\to \infty} \frac{1}{t}   {\rm Var}^{*} \left[ \int_0^t \alpha_{st} \, \rd \xi_s\right] = \lim_{t\to \infty} \frac{1}{t}   \int_0^t \psi''(\alpha_{st}) \alpha_{st}^2 \, \rd s
\end{eqnarray}
is finite, which can be seen using l'Hospital's rule. Hence, \eqref{Var} follows.
Now let us define
\begin{eqnarray}
  c(t) =   \mathbb{E}^{*} \left[\int_0^t \alpha_{st}\,  \rd \xi_s\right] - B(t, \omega t),
\end{eqnarray}
and recall from \eqref{help_lemma1_eq} that $c(t)$ grows linearly for large $t$. Let $t$ be large enough so that $c(t) >0$. We recall that if $X$ is a random variable such that ${\rm Var} \, [X] < \infty$, then for any constant $c > 0$ we have the Chebyshev inequality
\begin{equation}
\mathbb{P}\left[ | X - \mathbb{E}[X]| \geq c \right] \leq \frac{1}{c^2}  {\rm Var} \, [X] .
\end{equation}
In the present context it follows that
\begin{eqnarray}
   \mathbb{P}^{*} \left[ \left| \int_0^t \alpha_{st} \rd \xi_s -   \mathbb{E}^{*} \left[\int_0^t \alpha_{st} \rd \xi_s\right] \right| \geq c(t) \right] \leq \frac{1}{c(t)^2}{\rm Var}^{*} \left[ \int_0^t \alpha_{st} \rd \xi_s\right].
\end{eqnarray}
Since both $c(t)$ and the variance grow linearly in $t$ by \eqref{help_lemma1_eq}--\eqref{Var}, we see that
\begin{eqnarray}
  \lim_{t \to \infty}   \mathbb{P}^{*} \left[ \left| \int_0^t \alpha_{st} \rd \xi_s -   \mathbb{E}^{*} \left[\int_0^t \alpha_{st} \rd \xi_s\right] \right| \geq c(t) \right]  = 0. \label{Cheby}
\end{eqnarray}
For $t$ large enough so that $c(t) >0$, we have
\begin{eqnarray}
  \mathbb{E}^{*} \! \left[ \mathds{1} \left\{\int_0^t \alpha_{st} \rd \xi_s > B(t, \omega t) \right\} \right] \!\nonumber &=& \! \mathbb{P}^{*} \left[ \int_0^t \alpha_{st} \rd \xi_s > B(t, \omega t) \right] \\
  \nonumber  &\geq& \!   \mathbb{P}^{*} \left[ \left| \int_0^t \alpha_{st} \rd \xi_s -   \mathbb{E}^{*} \left[\int_0^t \alpha_{st} \rd \xi_s\right] \right| < c(t) \right] \\
  &=& \!1 - \mathbb{P}^{*} \left[ \left| \int_0^t \alpha_{st} \rd \xi_s -   \mathbb{E}^{*} \left[\int_0^t \alpha_{st} \rd \xi_s\right] \right| \geq c(t) \right] \!. 
\end{eqnarray}
Taking the limit on both sides as $t$ gets large, and using equation  \eqref{Cheby}, one finds that 
 \begin{eqnarray}
   \lim_{t \to \infty} \mathbb{E}^{*}\left[ \mathds{1} \left\{\int_0^t \alpha_{st} \rd \xi_s > B(t, \omega t) \right\} \right] =1. \label{step1}
  \end{eqnarray}
To proceed, recall from \eqref{PK_UI} and \eqref{exp_term} that in order to show that the pricing kernel is not UI when $R_\infty = 0$ one needs to prove that 
\begin{eqnarray}
   &&\lim_{\delta \to \infty} \sup_t  \bigg(P_{0t} \, \mathbb{E}^{*}\left[  {\mathds 1}\left\{ \int_0^t \alpha_{st} \, \rd \xi_s > B(t, \delta) \right\}\right]\bigg)  >0. \label{goal}
\end{eqnarray}
To see that  \eqref{goal} holds, note that
\begin{eqnarray}
	&& \lim_{\delta \to \infty} \sup_t  \bigg(P_{0t} \, \mathbb{E}^{*}\left[  {\mathds 1}\left\{ \int_0^t \alpha_{st} \, \rd \xi_s > B(t, \delta) \right\}\right]\bigg) \nonumber
	\\ &\quad& = \lim_{T \to \infty} \sup_t  \bigg(P_{0t} \, \mathbb{E}^{*}\left[  {\mathds 1}\left\{ \int_0^t \alpha_{st} \,  \rd \xi_s > B(t, \omega T) \right\}\right]\bigg) \nonumber
	\\
	&\quad& = \limsup_{T \to \infty} \, \sup_t \bigg(P_{0t} \, \mathbb{E}^{*}\left[  {\mathds 1}\left\{ \int_0^t \alpha_{st} \,  \rd \xi_s > B(t, \omega T) \right\}\right]\bigg) \nonumber 
	\\
	&\quad& \geq  \limsup_{T \to \infty} P_{0T} \, \mathbb{E}^{*}\left[  {\mathds 1}\left\{ \int_0^T \alpha_{sT}  \, \rd \xi_s > B(T, \omega T) \right\}\right] \nonumber 
	\\ &\quad& = \limsup_{T \to \infty} P_{0T}, 	\label{ineq_last}
\end{eqnarray}
where the final equality follows as a consequence of  equation \eqref{step1}. Keeping in mind that for the case under consideration we have $R_\infty = 0$, it follows from \eqref{logex} and \eqref{Lemma2_eq}  that $P_{0T}$ is of the form 
\begin{eqnarray}
P_{0T} =  \exp\left( -  \frac{1}{k} (r_0 -\theta) \left( 1 - \re^{-kT}\right) + f(T) \right)  \label{exp_factor}
\end{eqnarray}
for some function $f(T)$ that satisfies 
\begin{equation}
	\limsup_{T \to \infty} | f(T)| < \infty.
\end{equation}
We deduce that 
\begin{equation}
	\limsup_{T \to \infty}  P_{0T} =  \exp\left( -  \frac{1}{k} (r_0 -\theta) \right) 	\limsup_{T \to \infty} \, \exp\left( f(T) \right)  > 0.
\end{equation}
Therefore, the right hand side of \eqref{ineq_last} is strictly positive, and thus  we have shown that  \eqref{goal} holds, which concludes the proof. \hfill $\Box$

\section{Long-bond return in L\'evy-Vasicek model}
\noindent
We proceed to determine 
the return on a unit investment in the long bond in a L\'evy-Vasicek model.  Note that
\begin{eqnarray}
 \lim_{T \to \infty} \log \frac{P_{tT}}{P_{0T}}  = \lim_{T\to\infty} \left[- R_{tT} \,(T - t) + R_{0T}\,T \right] ,
\end{eqnarray}
from which it follows that 
\begin{eqnarray}
L_t &=&  
  (\theta + \psi(-\lambda))t + \frac{1}{k}(r_0 - r_t) - \lim_{T \to \infty} \int_0^t \psi(\alpha_{sT}) \rd s 
  \nonumber \\ &= & 
  (\theta + \psi(-\lambda))t + \frac{1}{k}(r_0 - r_t) - \psi\left(\frac{\sigma}{k} - \lambda \right)t \nonumber \\
  &=& R_{\infty} t + \frac{1}{k}(r_0 - r_t).
\end{eqnarray}
It is natural to ask whether it is possible to bring the long-bond return process into the form of a 
geometric price process. If we take into account relation (\ref{Rt_Levy}) for the integrated short rate and definition (\ref{eq:35}) 
for the excess rate of return, then after some algebra we deduce that 
\begin{eqnarray}
L_t = \exp \left[ \int_0^t \left( r_s  + R\left(\lambda, \frac{\sigma}{k}\right) \right)\rd s  + 
\frac{\sigma}{k} \xi_t  - \psi\left(\frac{\sigma}{k}\right) t \right].
\end{eqnarray} 
Thus, the form of the long-bond return is indeed that of a geometric asset price, with a 
L\'evy-Vasicek short rate $r_t$, risk aversion $\lambda$, and volatility $\sigma/k$. If we multiply this 
expression by the pricing kernel, then after some cancelation we obtain the geometric L\'evy 
martingale
\begin{eqnarray}
M_t  &=&
\exp \left[  \left(\frac{\sigma}{k} - \lambda\right) \xi_t -\psi\left(\frac{\sigma}{k} - \lambda\right) \,t \right] .
\end{eqnarray}
This gives the result that Ross recovery 
holds in a L\'evy-Vasicek model 
 if and only if \begin{eqnarray}
\lambda = \frac{\sigma}{k},
\end{eqnarray}
just as in the Brownian context. 
While it seems unlikely in practice that the interest-rate market price of risk is equal to the long-bond volatility, it would nevertheless be interesting to extend the analysis of Carr \& Yu (2012), Borovi\v{c}ka \emph{et al} (2014), Qin \emph{et al}  (2016), and others to the context of L\'evy models such as the ones developed here. 
It remains an open question whether a version of Ross recovery may yet survive if one works entirely in real terms.
What is clear, however, is that the matters investigated in the present paper, in particular the uniform integrability of pricing kernels and the nature of the long-bond return process, are important areas of investigation, and that their examination in varied contexts is likely to offer further insights into the behaviour of financial models over long time horizons. 

 \vspace{0.5cm} 

\begin{acknowledgments}
\noindent
We are grateful to participants at the Research in Options conference, B\'uzios, Rio de Janeiro (December 2014), the Brunel University Mathematical Finance Seminar (December 2014), the 4th WBS Interest Rate Conference, London (March 2015), and the 9th World Congress of the Bachelier Finance Society, New York (July 2016), where earlier versions of this work were presented, for helpful comments. This research was supported by a Brunel Research Initiative and Enterprise Fund Award.  Part of this work was completed at the Aspen Center for Physics, which is supported by National Science Foundation grant PHY-1066293. DCB acknowledges support from the Russian Science Foundation (project 16-11-10218).
\end{acknowledgments}

\end{document}